\documentclass[
  twocolumn,
  prl,
  amsmath,
  amssymb,
  superscriptaddress,
  floatfix
]{revtex4} 

\usepackage{bm}
\usepackage{dsfont}
\usepackage{graphicx}
\usepackage{pifont}
\usepackage{textcomp}
\usepackage{gensymb}
\usepackage{accents}
\usepackage{upgreek}
\usepackage{array}
\usepackage{tabularx}
\usepackage{color}
\usepackage{hhline}
\usepackage{multirow}
\usepackage{mathtools}
\usepackage{natbib}
\usepackage{hyperref}

\catcode`@=11

\def\nvphantom{\v@true\h@false\nph@nt}
\def\nhphantom{\v@false\h@true\nph@nt}
\def\nphantom{\v@true\h@true\nph@nt}
\def\nph@nt{\ifmmode\def\next{\mathpalette\nmathph@nt}%
  \else\let\next\nmakeph@nt\fi\next}
\def\nmakeph@nt#1{\setbox\z@\hbox{#1}\nfinph@nt}
\def\nmathph@nt#1#2{\setbox\z@\hbox{$\m@th#1{#2}$}\nfinph@nt}
\def\nfinph@nt{\setbox\tw@\null
  \ifv@ \ht\tw@\ht\z@ \dp\tw@\dp\z@\fi
  \ifh@ \wd\tw@-\wd\z@\fi \box\tw@}

\makeatletter
\def\NAT@bibsetnum#1{%
 \setlength{\topsep}{\z@}%
 \NATx@bibsetnum{#1}%
}%
\renewenvironment{thebibliography}[1]{%
 \NAT@thebibliography{#1}%
 \@clubpenalty\clubpenalty
 \let\@TBN@opr\present@bibnote
 \@FMN@list
}{%
 \@endnotesinbib
 \edef\@currentlabel{\arabic{NAT@ctr}}%
 \NAT@endthebibliography
 \global\let\auto@bib\@empty
}
\newcommand*{\supplementarystart}{%
  \close@column@grid%
  \clearpage%
  \onecolumngrid%
  \setcounter{equation}{0} 
  \setcounter{enumiv}{0} 
  \setcounter{figure}{0} 
  \setcounter{table}{0} 
  \setcounter{page}{1}
  \c@secnumdepth=4
  \renewcommand{\bibnumfmt}[1]{[s##1]} 
  \renewcommand{\@onlinecite}{
  \citealp} 
  \renewcommand{\cite}[1]{{[s}\onlinecite{##1}{]}}
  \renewcommand{\thefigure}{s\arabic{figure}}
  \renewcommand{\thetable}{s\Roman{table}}
  \renewcommand{\thepage}{s\arabic{page}}
  \renewcommand{\theequation}{s\arabic{equation}} 
}
\makeatother

\newcommand{\be}{\begin{equation}}
\newcommand{\e}{\end{equation}}
\newcommand{\beml}{\begin{subequations}}
\newcommand{\eml}{\end{subequations}}
\newcommand{\beq}{\begin{eqnarray}}
\newcommand{\eq}{\end{eqnarray}}
\newcommand{\ba}{\begin{array}}
\newcommand{\ea}{\end{array}}
\newcommand{\bpm}{\begin{pmatrix}}
\newcommand{\epm}{\end{pmatrix}}
\newcommand{\bc}{\begin{cases}}
\newcommand{\ec}{\end{cases}}

\newcommand{\bb}{\boldsymbol}

\renewcommand{\log}{\mathop{\mathrm{ln}}\nolimits}

\DeclareMathOperator{\tr}{Tr}

\DeclareMathOperator{\re}{Re}

\DeclareMathOperator{\sign}{sign}

\begin{document}

\title{Chiral ferromagnetism beyond Lifshitz invariants}

\author{I.\,A.~Ado}
\affiliation{Radboud University, Institute for Molecules and Materials, NL-6525 AJ Nijmegen, The Netherlands}

\author{A.~Qaiumzadeh}
\affiliation{Center for Quantum Spintronics, Department of Physics, Norwegian University of Science and Technology, NO-7491 Trondheim, Norway}

\author{A.~Brataas}
\affiliation{Center for Quantum Spintronics, Department of Physics, Norwegian University of Science and Technology, NO-7491 Trondheim, Norway}

\author{M.~Titov}
\affiliation{Radboud University, Institute for Molecules and Materials, NL-6525 AJ Nijmegen, The Netherlands}
\affiliation{ITMO University, Saint Petersburg 197101, Russia}

\begin{abstract}
We consider a contribution $w_{\text{ch}}$ to the micromagnetic energy density that is linear with respect to the first spatial derivatives of the local magnetization direction. For a generalized 2D Rashba ferromagnet, we present a microscopic analysis of this contribution and, in particular, demonstrate that it cannot be expressed through Lifshitz invariants beyond the linear order in the spin-orbit coupling (SOC) strength. Terms in $w_{\text{ch}}$ beyond Lifshitz invariants emerge as a result of spin rotation symmetry breaking caused by SOC. Effects of these terms on the phase diagram of magnetic states and spin-wave dispersion are discussed. Finally, we present a classification of terms in~$w_{\text{ch}}$, allowed by symmetry, for each crystallographic point group.
\end{abstract}

\maketitle
The Dzyaloshinskii-Moriya interaction (DMI)~\cite{Dzyaloshinsky1958, Moriya1960} is usually regarded as a key ingredient for the existence of chiral magnetism~\mbox{\cite{Bak1980, BogdanovYablonskii1989, Rossler2006, Bode2007, Tretiakov2010, Chen2013, Emori2013}}. In ferromagnets (FMs), DMI is described, in the continuum limit, by so-called Lifshitz invariants (LIs), antisymmetric combinations of the form
\be
\label{LI}
\mathcal L^{(k)}_{ij}=n_i\nabla_k n_j-n_j\nabla_k n_i,
\e
where $\bb n$ is a unit vector of the local magnetization direction~\cite{Bak1980, BogdanovYablonskii1989, Rossler2006, Tretiakov2010, Thiaville2012, Meynell2014, Freimuth2017}. In a broader sense, one can consider a ``general chiral contribution''
\be
\label{DMI_general}
w_{\text{ch}}=
\sum_{\beta\gamma}{\Omega^{\text{ch}}_{\beta\gamma}\,\nabla_\beta\, n_\gamma}
\e
to the micromagnetic energy density that is linear with respect to the first spatial derivatives of $\bb n$, but is not necessarily expressed only in terms of LIs. Below, we refer to $w_{\text{ch}}$ as the chiral energy density.

Time-reversal symmetry dictates that elements of the tensor $\Omega^{\text{ch}}$ should be odd with respect to a transformation $\bb n\to-\bb n$~\cite{Landafshitz}. Usually, it is simply assumed that $\Omega^{\text{ch}}_{\beta\gamma}(\bb n)$ are linear functions of the components $n_i$. In this case, $w_{\text{ch}}$ reduces to a linear combination of LIs and the corresponding symmetric terms $\nabla_k(n_i n_j)$~\cite{Hals2017, Hals2019}. The latter describe only the effects of boundaries~\cite{Dzyaloshinskii1964}.

Quite recently, such boundary effects came into the focus of phenomenological studies in systems with the $C_{\infty v}$ point group symmetry. The authors of Refs.~\cite{Hals2017, Hals2018} demonstrated that the terms $\nabla_k(n_i n_j)$ in $w_{\text{ch}}$ may become important in thin film systems. In particular, it was suggested that such terms can lead to the formation of magnetic twist states~\cite{Hals2017} and contribute to the stability of skyrmions~\cite{Hals2018}.

In this paper, the chiral energy density is addressed beyond the assumption of the linear dependence of $\Omega^{\text{ch}}_{\beta\gamma}(\bb n)$ on $n_i$. Both microscopically and phenomenologically, we demonstrate that LIs can be insufficient for describing the chirality of a ferromagnet in the continuum limit, even in the absence of boundary effects (e.g., when the system is effectively infinite).

Let us start with a microscopic analysis of $w_{\text{ch}}$ for a particular 2D model system with the $C_{\infty v}$ symmetry. We consider a FM layer coupled to a 2DEG with spin-orbit coupling (SOC) of Rashba type and assume that the 2DEG is described by the Hamiltonian
\be
\label{Hamiltonian}
\mathcal{H}=\xi(p)+\alpha_{\text{\tiny R}}\zeta(p)\, [\bb{p}\times\bb{\sigma}]_z  + J_{\text{sd}} S\,\bb n(\bb r)\cdot\bb{\sigma},
\e
where the term $\xi(p)$ parametrizes the nonrelativistic electron dispersion, while the function $\zeta(p)$ quantifies the momentum-dependent \hspace{-0.16ex}Rashba \hspace{-0.16ex}SOC \hspace{-0.16ex}of \hspace{-0.16ex}strength \hspace{-0.16ex}$\alpha_{\text{\tiny R}}$. \hspace{-0.23ex}In the last term of Eq.~(\ref{Hamiltonian}), $\bb\sigma$ stands for the vector of Pauli matrices, while $J_{\textrm{sd}}$ represents the strength of the $s$-$d$-type exchange interaction between the 2DEG and localized FM spins of the absolute value $S$.

Using the model of Eq.~(\ref{Hamiltonian}), $w_{\text{ch}}$ has been computed recently~\cite{linearDMI_us} in the lowest (linear) order with respect to~$\alpha_{\text{\tiny R}}$, with the result
\be
\label{DMI_energy_linear}
w_{\text{ch}}=-D\left(\mathcal{L}_{zx}^{(x)}-\mathcal{L}_{yz}^{(y)}\right)=D\,\bb n \cdot [[\bb e_z\times\bb\nabla]\times\bb n],
\e
where $\bb\nabla=(\nabla_x,\nabla_y)$ and $D$ is a DMI constant proportional to~$\alpha_{\text{\tiny R}}$. We are about to show that, beyond the linear order in the SOC strength, Eq.~(\ref{DMI_energy_linear}) transforms into
\begin{multline}
\label{DMI_energy_general}
w_{\text{ch}}=D_\parallel(n_z^2)\,\bb n \cdot [[\bb e_z\times\bb\nabla]\times\bb n_\parallel]\\+D_\perp(n_z^2)\,\bb n \cdot [[\bb e_z\times\bb\nabla]\times\bb n_\perp],
\end{multline}
where $D_\parallel$ differs from $D_\perp$, and $\bb n_{\parallel/\perp}$ denotes the in-plane/perpendicular-to-the-plane component of $\bb n$,
\be
\label{components_of_n}
\bb n=\bb n_{\parallel}+\bb n_{\perp}, \qquad \bb n_{\perp}=\bb e_z n_z=\bb e_z \cos{\theta}.
\e
Note that the right hand side of Eq.~(\ref{DMI_energy_general}) is no longer expressed in terms of LIs, as one can deduce from a direct expansion of the vector products.

In order to derive Eq.~(\ref{DMI_energy_general}), we use a general expression for the tensor $\Omega^{\text{ch}}$~\cite{linearDMI_us}:
\begin{multline}
\label{DMI_tensor}
\phantom{\Biggr|}
\Omega^{\text{ch}}_{\beta\gamma}=T\frac{J_{\text{sd}} S}{2\pi\hbar}
\re
\int d \varepsilon\,
g(\varepsilon)
\int
\frac{d^2 p}{(2\pi)^2}\\
\times\tr{\Bigl(
G^{R}\sigma_\gamma\,G^{R}\,v_\beta\,G^{R}-G^{R}\,v_\beta\,G^{R}\sigma_\gamma\,G^{R}
\Bigr)},
\end{multline}
where $\bb v=\partial \mathcal H/\partial \bb p$ is the velocity operator, the retarded Green's function $G^R$ describes a system with homogeneous magnetization, and $\tr$ stands for the matrix trace operation. In Eq.~(\ref{DMI_tensor}), we also use the notation $g(\varepsilon)=\log{\left(1+\exp{\left[(\mu-\varepsilon)/T\right]}\right)}$, where $\mu$ and $T$ are the chemical potential and temperature, respectively. The Green's function $G^R$, in the momentum representation, takes the form 
\be
\label{green's_function}
G^R=\frac{\varepsilon-\xi(p)+\alpha_{\text{\tiny R}}\zeta(p)\, [\bb{p}\times\bb{\sigma}]_z  + J_{\text{sd}}S\,\bb n\cdot\bb{\sigma}}{(\varepsilon-\varepsilon_+(\bb p)+ i0)(\varepsilon-\varepsilon_-(\bb p)+ i0)},
\e
where the spectral branches $\varepsilon_{\pm}(\bb p)=\xi(p)\pm\Delta(\bb p)$ are parameterized by
\begin{gather*}
\Delta(\bb p)=\sqrt{\Delta_{\text{sd}}^2+[\alpha_{\text{\tiny R}}p\,\zeta(p)]^2-2\alpha_{\text{\tiny R}}\varsigma\Delta_{\text{sd}}\, p\,\zeta(p)\sin{\theta}\sin{\varphi}},\\
\Delta_{\text{sd}}=\vert J_{\text{sd}}\vert S, \quad \varsigma=\sign{J_{\text{sd}}}.
\end{gather*}
Here, $\theta$ stands for the polar angle of~$\bb n$ with respect to the $z$ axis, while $\varphi$ is the angle between the momentum~$\bb p$ and the in-plane component $\bb n_\parallel$ of the vector~$\bb n$.

Substitution of Eq.~(\ref{green's_function}) into Eq.~(\ref{DMI_tensor}) followed by the matrix trace calculation and integration over~$\varepsilon$ produces the following outcome:
\begin{multline}
\label{calculated_DMI_tensor}
\Omega^{\text{ch}}_{\beta\gamma}=W n_{\gamma}n_{\beta}
+D_{\parallel}(1-\delta_{\gamma z})
\sum\limits_{i j}{n_i\epsilon_{i j\gamma}\epsilon_{j z \beta}}
\\
+D_{\perp}\delta_{\gamma z}
\sum\limits_{i j}{n_i\epsilon_{i j\gamma}\epsilon_{j z \beta}},
\end{multline}
where $\epsilon_{q_1 q_2 q_3}$ denotes the three-dimensional Levi-Civita symbol, and $\delta_{q_1 q_2}$ is the Kronecker delta. The functions $D_\parallel$ and $D_\perp$ can be expressed as
\be
\label{D's_general}
D_a=\frac{\alpha_{\text{\tiny R}}\Delta_{\text{sd}}^2 T}{2\hbar}
\int{\frac{d^2 p}{(2\pi)^2}\,\mathcal D_a(\bb p)\left(\frac{g_+-g_-}{[\Delta(\bb p)]^3}-\frac{g_+' + g_-'}{[\Delta(\bb p)]^2}\right)},
\e
where $a=\,\parallel,\perp$ and we use the notations
\begin{gather}
\label{D's_micro_1}
g_\pm=g(\varepsilon_\pm(\bb p)), \quad g_\pm'=\partial g/\partial\varepsilon\bigr\vert_{\varepsilon=\varepsilon_\pm(\bb p)},
\\
\label{D's_micro_2}
\mathcal D_\parallel(\bb p)=\zeta(p)+p\,\zeta'(p)\sin^2{\varphi},
\\
\label{D's_micro_3}
\mathcal D_\perp(\bb p)=\mathcal D_\parallel(\bb p)+\frac{p\,\zeta'(p)\cos{2\varphi}}{\sin^2{\theta}}-\frac{\alpha_{\text{\tiny R}}p\,\zeta^2(p)\sin{\varphi}}{\varsigma\Delta_{\text{sd}}\sin{\theta}},
\end{gather}
with $\zeta'(p)=\partial \zeta/\partial p$. 

To translate Eq.~(\ref{calculated_DMI_tensor}) into the expression for $w_{\text{ch}}$, we first note that the value of $W$ is totally irrelevant for the final result. Indeed, upon substitution of Eq.~(\ref{calculated_DMI_tensor}) into Eq.~(\ref{DMI_general}), the first term on the right hand side of Eq.~(\ref{calculated_DMI_tensor}) produces a contribution that is equal to $(W/2)(\bb n_\parallel\cdot\bb\nabla)\bb n^2$. Due to the constraint $\bb n^2\equiv1$, it vanishes. The remaining two terms in Eq.~(\ref{calculated_DMI_tensor}) correspond to the double vector products in Eq.~(\ref{DMI_energy_general}). Noting that the dependence of $D_\parallel$ and $D_\perp$ on the vector $\bb n$, in the highly symmetric model of Eq.~(\ref{Hamiltonian}), can be expressed as $D_a=D_a(n_z^2)$~\cite{supplementary}, we therefore conclude the microscopic derivation of Eq.~(\ref{DMI_energy_general}).

The fact that $D_\parallel$ and $D_\perp$ both turn out to be functions of $\bb n$ has quite a few important consequences. Ignoring, for\hspace{-0.25ex} a\hspace{-0.25ex} moment,\hspace{-0.25ex} microscopic\hspace{-0.25ex} details,\hspace{-0.25ex} let\hspace{-0.25ex} us\hspace{-0.25ex} rewrite\hspace{-0.25ex} Eq.~(\ref{DMI_energy_general})\hspace{-0.25ex} as
\be
\label{DMI_energy_general_2}
w_{\text{ch}}=D_\perp(\bb n_\parallel \cdot \bb \nabla) n_z-D_\parallel n_z(\bb\nabla \cdot\bb n_\parallel).
\e
Integration over space defines the total micromagnetic chiral energy $\mathcal W_{\text{ch}}=\int{dx\, dy\,w_{\text{ch}}}$. Performing integration by parts and disregarding contributions from the boundaries, we obtain a different representation of the density,
\begin{multline}
\label{DMI_energy_general_3}
w_{\text{ch}}=-D_\perp n_z(\bb\nabla \cdot\bb n_\parallel)+D_\parallel(\bb n_\parallel \cdot \bb \nabla) n_z
\\
-n_z\frac{\partial D_\perp}{\partial\theta}(\bb n_\parallel \cdot \bb \nabla)\theta
+n_z\frac{\partial D_\parallel}{\partial\theta}(\bb n_\parallel \cdot \bb \nabla)\theta,
\end{multline}
where we have taken into account that the spatial dependence of $D_a=D_a(\cos^2{\theta})$ originates solely from the spatial dependence of the polar angle $\theta=\theta(\bb r)$. By taking the half sum of Eqs.~(\ref{DMI_energy_general_2})~and~(\ref{DMI_energy_general_3}), we arrive at the result
\begin{gather}
\label{DMI_energy_linear_novel}
w_{\text{ch}}=-D_{\text{as}}\left(\mathcal{L}_{zx}^{(x)}-\mathcal{L}_{yz}^{(y)}\right)+D_\text{diff}\,n_z(\bb n_\parallel \cdot \bb \nabla)\theta,\\
\label{D_diff_and_D_as_def}
D_{\text{as}}=\frac{D_\parallel+D_\perp}{2}, \qquad
D_{\text{diff}}=\frac{\partial}{\partial\theta}\frac{D_\parallel-D_\perp}{2},
\end{gather}
that demonstrates an essential separation of $w_{\text{ch}}$ into LI-type contributions and contributions of a different symmetry. 

The first term on the right hand side of Eq.~(\ref{DMI_energy_linear_novel}) has the structure of the DMI energy density for a system of the $C_{\infty v}$ class, Eq.~(\ref{DMI_energy_linear}). The second term, however, displays a non-LI-type symmetry and therefore does not originate from DMI. Importantly, it cannot be ``integrated out'' by means of a partial integration as opposed to the ``boundary terms'' $\nabla_k(n_i n_j)$. We explicitly note that $D_\text{diff}\neq 0$ requires at least one of the functions, $D_\parallel$ and $D_\perp$, to depend on $\bb n$, which is possible due to broken spin rotation symmetry. Similar orientational anisotropy of SOC-related phenomena has been observed recently                                                                                                                                                                                                                                                                                                                                                                                                                                                                                                                                                                                                                                                                                                                                                                                                                                                                                                                                                                                                           ~\cite{crazySOC(Kerr),crazySOC-1(GD),crazySOC-2(GD)}.

Remarkably, the phase diagram of magnetic states is affected by $D_\text{diff}$ as well as by $D_\text{as}$. Both functions incorporate an infinite amount of Fourier harmonics,
\begin{alignat}{4}
\label{D_as_Fourier}
&D_{\text{as}}&\,=\,&D^{(0)}_{\text{as}}&\,+\,&D^{(2)}_{\text{as}}\cos{2\theta}&\,+\,&D^{(4)}_{\text{as}}\cos{4\theta}+\dots,
\\
\label{D_diff_Fourier}
&&&D_{\text{diff}}&\,=\,&D^{(2)}_{\text{diff}}\sin{2\theta}&\,+\,&D^{(4)}_{\text{diff}}\sin{4\theta}+\dots,
\end{alignat}
which, obviously, complicates the minimization of the micromagnetic energy functional. Nevertheless, the role of the term $D_\text{diff}\,n_z(\bb n_\parallel \cdot \bb \nabla)\theta$ can be illustrated by using a simple example. Let us consider a domain wall (DW)
\be
\label{DW}
\theta\bigr\vert_{x\to -\infty}=\pi, \quad \theta\bigr\vert_{x\to +\infty}=0, \quad \phi\equiv\phi_0
\e
with the fixed azimuthal angle $\phi$ of the vector $\bb n=\bb n(x)$. Assuming the DW size in the $y$ direction to be equal to~$L$, we can compute the corresponding total chiral energy $\mathcal W^{\text{DW}}_{\text{ch}}=L\int{dx\,w_{\text{ch}}}$ from Eqs.~(\ref{DMI_energy_linear_novel}),~(\ref{D_as_Fourier}), and (\ref{D_diff_Fourier}). Making use of the relation $(\nabla_x\theta)dx=d\theta$ to reduce the integration over $x$ to the integration over $\theta$, we find
\begin{multline}
\label{DW_energy}
\mathcal W^{\text{DW}}_{\text{ch}}=-L\cos{\phi_0}\int_\pi^0{d\theta\,\Bigl(D_{\text{as}}-D_{\text{diff}}\sin{\theta}\cos{\theta}\Bigr)}
\\=
\pi L\cos{\phi_0}\left(D^{(0)}_{\text{as}}-\frac{1}{4}D^{(2)}_{\text{diff}}\right),
\end{multline}
where the orthogonality of the sine functions has been taken into account. Note that this result is independent of the particular shape of the DW profile $\theta(x)$.

The contribution to $\mathcal W^{\text{DW}}_{\text{ch}}$ that originates from the\vspace{-0.32ex} ``antisymmetric part'' $-D_{\text{as}}\bigl(\mathcal{L}_{zx}^{(x)}-\mathcal{L}_{yz}^{(y)}\bigr)$ of the chiral energy density has been computed before (see, e.\,g., Eq.~(19) in Ref.~\cite{Wakatsuki2015TI}). The second contribution provided by the term $D_\text{diff}\,n_z(\bb n_\parallel \cdot \bb \nabla)\theta$ is the novel result of this paper. As can be seen from Eq.~(\ref{DW_energy}), \vspace{-0.32ex}the DW chiral energy depends equally on the $D^{(0)}_{\text{as}}$ and $D^{(2)}_{\text{diff}}$ Fourier harmonics. Thus, indeed, the chirality of a ferromagnet in general cannot be properly analysed (in the continuum limit) without consideration of the non-LI-type contributions to $w_{\text{ch}}$. Despite being simplified, the ansatz of Eq.~(\ref{DW}) serves as a good illustration of the importance of such contributions. We certainly expect them to be relevant for more complex structures~\cite{Tretiakov2010} as well.

To perform an analogous to Eq.~(\ref{DMI_energy_linear_novel}) separation of $w_{\text{ch}}$ into LI-type and non-LI-type terms for an arbitrary FM, we assume that
\be
\label{chiral_energy_gen}
w_{\text{ch}}=\sum_{ijk}{D_{ijk}\,n_i\nabla_k n_j},
\e
where $D_{ijk}$ are even functions of $\bb n$. Below, we refer to the tensor with the components $D_{ijk}$ as the chiral tensor. Symmetrization of Eq.~(\ref{chiral_energy_gen}) gives~\cite{Hals2017, Hals2019}
\be
w_{\text{ch}}=\frac{1}{2}\sum_{ijk}{\left[D_{ijk}^{\text{as}}\mathcal L^{(k)}_{ij}+D_{ijk}^{\text{sym}}\nabla_k\left(n_i n_j\right)\right]},
\e
with $D_{ijk}^{\text{as(sym)}}=(D_{ijk}\mp D_{jik})/2$. Applying integration by parts to the second term inside the brackets and disregarding contributions from the boundaries, we obtain
\begin{gather}
\nonumber
w_{\text{ch}}=\frac{1}{2}\sum_{ijk}{\left[D_{ijk}^{\text{as}}\mathcal L^{(k)}_{ij}-\frac{\partial D_{ijk}^{\text{sym}}}{\partial \theta}\Theta^{(k)}_{ij}-\frac{\partial D_{ijk}^{\text{sym}}}{\partial \phi}\Phi^{(k)}_{ij}\right]},\\
\label{theta_and_phi_def}
\Theta^{(k)}_{ij}=n_i n_j\nabla_k\theta, \qquad \Phi^{(k)}_{ij}=n_i n_j\nabla_k\phi,
\end{gather}
where $\theta$ and $\phi$ are, as before, the polar and azimuthal angles of $\bb n$, respectively.

Using a standard \vspace{-0.42ex}symmetry analysis~\cite{Authier2003book, Hals2017}, one can identify the coefficients $\mathcal L^{(k)}_{ij}$, $\Theta^{(k)}_{ij}$, $\Phi^{(k)}_{ij}$ that are allowed in $w_{\text{ch}}$ by a point group symmetry of a particular system. The corresponding results for all crystallographic point groups (except $C_1$, $C_{1v}$, and $C_{1h}$ that we address in the Supplemental Material~\cite{supplementary}) are collected in Table~\ref{table}. Remarkably, for the classes $C_{3h}$, $D_{3h}$, and $T_d$, DMI does not contribute to the chiral energy density. Non-LI-type terms are the only source of chirality in FMs described by these three groups.

Let us now use the results of Eqs.~(\ref{D's_general}),~(\ref{D's_micro_1}),~(\ref{D's_micro_2}),~(\ref{D's_micro_3}) and return to the microscopic analysis of the functions $D_\parallel$ and $D_\perp$ for the generalized Rashba model of Eq.~(\ref{Hamiltonian}). First, it is easy to observe that, in the leading (linear) order with respect to small $\alpha_{\text{\tiny R}}$, the angle integration in Eq.~(\ref{D's_general}) can be performed straightforwardly. This leads to the result $D_\parallel=D_\perp=D_{\text{as}}=D$, where $D$ is the DMI constant given by Eq.~(5) of Ref.~\cite{linearDMI_us}.

In the other limit, $\Delta_{\text{sd}}\to 0$, the coefficients of the chiral tensor do not coincide: one generally finds $D_\parallel\neq D_\perp$, in the leading (second) order with respect to small~$\Delta_{\text{sd}}$. Nevertheless, in this case, the quantity $D_\parallel-D_\perp$ turns out to be independent of $\theta$ and hence $D_{\text{diff}}=0$~\cite{supplementary}. Therefore, for either weak SOC or weak $s$-$d$ exchange, the chiral energy density can be described by LIs alone, at least as long as the boundary effects are disregarded. Any possible effect of $D_{\text{diff}}$ is absent in these two limits.

A further asymptotic analysis~\cite{supplementary} shows that the function $D_{\text{diff}}$ does not contain contributions in the order $\alpha_{\text{\tiny R}}^3$. Indeed, in this order, $D_\parallel\neq D_\perp$ -- yet, again, the difference between $D_\parallel$ and $D_\perp$ does not depend on~$\theta$. In general, the expansions of $D_{\text{diff}}$ in small $\alpha_{\text{\tiny R}}$ and small $\Delta_{\text{sd}}$ start with the contributions of the order~$\alpha_{\text{\tiny R}}^5$ and~$\Delta_{\text{sd}}^4$, respectively. Moreover, for the leading-order asymptotics, only the\vspace{-0.32ex} first Fourier harmonic is nonvanishing, so that $D_{\text{diff}}=D^{(2)}_{\text{diff}}\sin{2\theta}$.

\begin{figure}[t!]
\includegraphics[width=0.99\columnwidth]{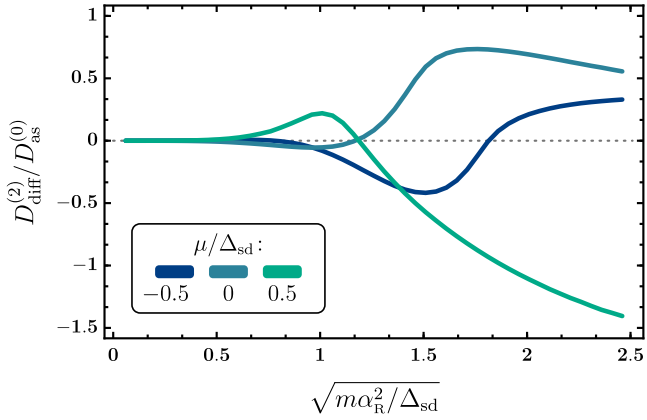}
\caption{The ratio between the leading Fourier coefficients of the functions $D_{\text{diff}}$ and $D_{\text{as}}$ in the Bychkov-Rashba model. All three curves are obtained numerically by changing the parameter $\alpha_{\text{\tiny R}}$ (with others fixed). Temperature is set to zero.}
\label{fig::D_diff}
\end{figure}

For the particular Bychkov-Rashba model~\cite{BychkovRashba1984} characterized by the choice $\xi(p)=p^2/2m$ and $\zeta(p)\equiv 1$ in Eq.~(\ref{Hamiltonian}), we find at $T=0$ for the first nonzero terms of the expansions (in $\alpha_{\text{\tiny R}}$ and $\Delta_{\text{sd}}$, respectively):
\begin{align*}
D^{(2)}_{\text{diff}}&=-
\frac{m \alpha_{\text{\tiny R}}\Delta_{\text{sd}}}{128\pi\hbar}
\left(\frac{m\alpha_{\text{\tiny R}}^2}{\Delta_{\text{sd}}}\right)^2
\begin{cases}
Q\,(\mu/\Delta_{\text{sd}}),
&\vert\mu\vert<\Delta_{\text{sd}}\\
0, &\phantom{\vert}\mu\phantom{\vert}>\Delta_{\text{sd}}
\end{cases},
\\
D^{(2)}_{\text{diff}}&=-
\frac{m \alpha_{\text{\tiny R}}\Delta_{\text{sd}}}{16\pi\hbar}
\left(\frac{\Delta_{\text{sd}}}{m\alpha_{\text{\tiny R}}^2}\right)^3
\begin{cases}
R\,(\mu/m\alpha_{\text{\tiny R}}^2), &\mu<0\\
0, &\mu>0\\
\end{cases},
\end{align*}
where we have introduced $Q(x)=35 x^4-30 x^2+3$ and $R(x)=(35 x^2+40 x+12)/(1+2x)^{5/2}$, with $R(x)\equiv 0$ for $x<-1/2$. In principle, it is clear that, in this simple model, $D_{\text{diff}}$ is determined by three independent energy scales: $m\alpha_{\text{\tiny R}}^2$, $\Delta_{\text{sd}}$, and $\mu$. Intuitively, one would expect $\vert D_{\text{diff}}\vert$ to be maximal when $m\alpha_{\text{\tiny R}}^2$ and $\Delta_{\text{sd}}$ are of a comparable magnitude. Our perturbative analysis agrees with this conjecture. We also illustrate the\vspace{-0.32ex} latter in Fig.~\ref{fig::D_diff}, by plotting the ratio $D^{(2)}_{\text{diff}}/D^{(0)}_{\text{as}}$ as a function of the SOC strength. The absolute value of this ratio, and even its sign, are sensitive to variation of the chemical potential. One might recognize this as a possibility to gain additional means of magnetic order tuning by means of gate voltage control.

Notably, in the Bychkov-Rashba model, the leading-order asymptotics of $D_{\text{diff}}$ vanish at zero temperature when both spin subbands are partly occupied. This is not accidental. In fact, chiral terms in the micromagnetic energy density (including those originating from DMI) are totally absent in this case, $D_\parallel\equiv 0$ and $D_\perp\equiv 0$, regardless of the values of $\alpha_{\text{\tiny R}}$ and $\Delta_{\text{sd}}$~\cite{supplementary}. Such a peculiarity, however, is a property of the specific model and does not characterize the symmetry class ($C_{\infty v}$) to which the latter corresponds. Indeed, one may consider a slightly more general example, with $\xi(p)=(p^2/2m)/\left(1+\kappa\, p^2/2m\right)$ and $\zeta(p)=1/\left(1+\lambda\, p^2/2m\right)$, where the positive parameters $\kappa$ and $\lambda$ represent deviations from parabolic band dispersion~\cite{nonparabolicity-I,nonparabolicity-II,nonparabolicity-III} and nonlinear dependence of the Rashba SOC on momentum~\cite{nonlinear_RSOC_theory,nonlinear_RSOC_DFT,nonlinear_RSOC_exp}, respectively. In this model, finite $w_{\text{ch}}$ for two partly occupied subbands is restored. In particular, for $\mu>\Delta_{\text{sd}}$, we find a surprisingly compact result
in the leading $\alpha_{\text{\tiny R}}^5$ order,
\begin{equation}
\label{Ddiff_eta_lambda}
D^{(2)}_{\text{diff}}=
-\frac{14m \alpha_{\text{\tiny R}}\Delta_{\text{sd}}}{3\pi\hbar}
\left(m\alpha_{\text{\tiny R}}^2\right)^2\Delta_{\text{sd}}^3\,(\kappa-\lambda)^4(4\kappa-9\lambda),
\end{equation}
where the temperature is set to zero and $\kappa\approx\lambda$ are both considered small in comparison with $\mu^{-1}$ and $\Delta_{\text{sd}}^{-1}$.

\begin{figure}[b!]
\includegraphics[width=0.99\columnwidth]{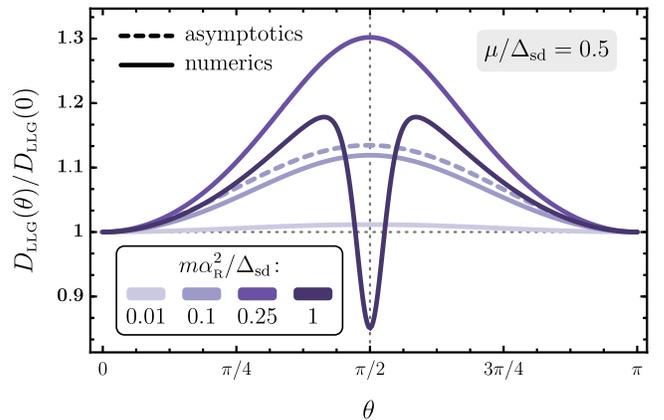}
\caption{The quantity $D_{\text{\tiny{LLG}}}$ in the Bychkov-Rashba model as a function of the polar angle of magnetization direction at zero temperature. Solid curves represent numerical results. For \mbox{$m\alpha_{\text{\tiny R}}^2/\Delta_{\text{sd}}=0.1$}, the asymptotic expansion up to the order~$\alpha_{\text{\tiny R}}^3$ (given by Eq.~(s26) in the Supplemental Material~\cite{supplementary}) is shown for comparison. For \mbox{$m\alpha_{\text{\tiny R}}^2/\Delta_{\text{sd}}=0.01$}, numerical and asymptotic curves are indistinguishable.}
\label{fig::D_LLG}
\end{figure}

In the final part of the paper, we briefly discuss how the chiral energy density with the symmetry of Eq.~(\ref{DMI_energy_general}) affects spin-wave dispersion. The effective field arising due to $w_{\text{ch}}$ is proportional to the functional derivative $\delta\mathcal W_{\text{ch}}/\delta\boldsymbol n$. Taking advantage of the fact that both $D_\parallel$ and $D_\perp$ can be considered independent of $\bb n_\parallel=(n_x, n_y)$, we find $\delta\mathcal W_{\text{ch}}/\delta\boldsymbol n=2D_{\text{\tiny{LLG}}} \bb u$, where
\begin{gather}
\label{D_LLG}
D_{\text{\tiny{LLG}}}=\frac{1}{2}\left(D_\parallel+D_\perp+n_z\frac{\partial D_\parallel}{\partial n_z}\right),
\\
\boldsymbol u=\bb\nabla n_z-\boldsymbol e_z(\bb\nabla\cdot\bb n_\parallel).
\end{gather}
The corresponding contribution to the Landau-Lifshitz-Gilbert (LLG) equation shifts the frequency of a spin wave by a term linear in wave vector $\bb k$~\cite{Moon2013linear_shift, Di2015D_LLG, Belmeguenai2015linear_shiftBLS}. Importantly, the frequency difference $\Delta f$ between spin waves with wave vectors $\bb k$ and $-\bb k$ is experimentally measurable~\cite{Zakeri2010, Di2015D_LLG, Belmeguenai2015linear_shiftBLS, Nembach2015BLS}. In the present case, such difference should equal
\be
\label{frequency}
\Delta f=\frac{2\gamma D_{\text{\tiny{LLG}}}}{\pi M_s}\left[\bb n\times \bb k\right]_z,
\e
where $M_s$ is a saturation magnetization and $\gamma$ denotes the gyromagnetic ratio.

\begingroup
\begin{table*}
\begin{center}
{\renewcommand{\arraystretch}{1.35}
\tabcolsep=2.4pt
\begin{tabular}{c|c|c}
symmetry & LI-type terms & non-LI-type terms \\
\hline
$C_2$ ($D_1$)
& $\mathcal L^{(x)}_{zx}$;\nphantom{;}\phantom{+} $\mathcal L^{(y)}_{yz}$;\nphantom{;}\phantom{+} $\mathcal L^{(x)}_{yz}$;\nphantom{;}\phantom{+} $\mathcal L^{(y)}_{zx}$;\nphantom{;}\phantom{+} $\mathcal L^{(z)}_{xy}$
& \phantom{$\mathcal Q$;\nphantom{;}\phantom{+} $\mathcal S$;\nphantom{;}\phantom{+}} $\mathcal A^{(x)}_{zx}$;\nphantom{;}\phantom{+} $\mathcal A^{(y)}_{yz}$;\nphantom{;}\phantom{+} $\mathcal A^{(x)}_{yz}$;\nphantom{;}\phantom{+} $\mathcal A^{(y)}_{zx}$;\nphantom{;}\phantom{+} $\mathcal A^{(z)}_{xy}$;\nphantom{;}\phantom{+} $\mathcal A^{(z)}_{xx}$;\nphantom{;}\phantom{+} $\mathcal A^{(z)}_{yy}$ \\
\hline
$C_{2v}$ ($D_{1h}$)
& $\mathcal L^{(x)}_{zx}$;\nphantom{;}\phantom{+} $\mathcal L^{(y)}_{yz}$\phantom{;\nphantom{;}\phantom{+} $\mathcal L^{(x)}_{yz}$;\nphantom{;}\phantom{+} $\mathcal L^{(y)}_{zx}$;\nphantom{;}\phantom{+} $\mathcal L^{(z)}_{xy}$}
& \phantom{$\mathcal Q$;\nphantom{;}\phantom{+} $\mathcal S$;\nphantom{;}\phantom{+}} $\mathcal A^{(x)}_{zx}$;\nphantom{;}\phantom{+} $\mathcal A^{(y)}_{yz}$;\phantom{\nphantom{;}\phantom{+} $\mathcal A^{(x)}_{yz}$;\nphantom{;}\phantom{+} $\mathcal A^{(y)}_{zx}$;\nphantom{;}\phantom{+} $\mathcal A^{(z)}_{xy}$;\nphantom{;}\phantom{+}} $\mathcal A^{(z)}_{xx}$;\nphantom{;}\phantom{+} $\mathcal A^{(z)}_{yy}$ \\
\hline
$D_2$
& \phantom{$\mathcal L^{(x)}_{zx}$;\nphantom{;}\phantom{+} $\mathcal L^{(y)}_{yz}$;}\nphantom{;}\phantom{+} $\mathcal L^{(x)}_{yz}$;\nphantom{;}\phantom{+} $\mathcal L^{(y)}_{zx}$;\nphantom{;}\phantom{+} $\mathcal L^{(z)}_{xy}$
& \phantom{$\mathcal Q$;\nphantom{;}\phantom{+} $\mathcal S$;\nphantom{;}\phantom{+}} \phantom{$\mathcal A^{(x)}_{zx}$;\nphantom{;}\phantom{+} $\mathcal A^{(y)}_{yz}$;}\nphantom{;}\phantom{+} $\mathcal A^{(x)}_{yz}$;\nphantom{;}\phantom{+} $\mathcal A^{(y)}_{zx}$;\nphantom{;}\phantom{+} $\mathcal A^{(z)}_{xy}$\phantom{;\nphantom{;}\phantom{+} $\mathcal A^{(z)}_{xx}$;\nphantom{;}\phantom{+} $\mathcal A^{(z)}_{yy}$} \\
\hline
$D_{2d}$
& \phantom{$\mathcal L^{(x)}_{zx}+\mathcal L^{(y)}_{yz}$;}\nphantom{;}\phantom{+} $\mathcal L^{(x)}_{yz}-\mathcal L^{(y)}_{zx}$\phantom{+ $\mathcal L^{(z)}_{xy}$}
& \phantom{$\mathcal Q$;\nphantom{;}\phantom{+} $\mathcal S$;\nphantom{;}\phantom{+}} \phantom{$\mathcal A^{(x)}_{zx}$;\nphantom{;}\phantom{+} $\mathcal A^{(y)}_{yz}$;}\nphantom{;}\phantom{+} $\mathcal A^{(x)}_{yz}+\mathcal A^{(y)}_{zx}$;\nphantom{;}\phantom{+} $\mathcal A^{(z)}_{xy}$\phantom{;\nphantom{;}\phantom{+} $\mathcal A^{(z)}_{xx}$;\nphantom{;}\phantom{+} $\mathcal A^{(z)}_{yy}$} \\
\hline
$C_3$
& $\mathcal L^{(x)}_{zx}-\mathcal L^{(y)}_{yz}$;\nphantom{;}\phantom{+} $\mathcal L^{(x)}_{yz}+\mathcal L^{(y)}_{zx}$;\nphantom{;}\phantom{+} $\mathcal L^{(z)}_{xy}$
& $\mathcal Q$;\nphantom{;}\phantom{+} $\mathcal S$;\nphantom{;}\phantom{+} $\mathcal A^{(x)}_{zx}+\mathcal A^{(y)}_{yz}$;\nphantom{;}\phantom{+} $\mathcal A^{(x)}_{yz}-\mathcal A^{(y)}_{zx}$;\nphantom{;}\phantom{+} \phantom{$\mathcal A^{(z)}_{xy}$;}\nphantom{;}\phantom{+} $\mathcal A^{(z)}_{xx}+\mathcal A^{(z)}_{yy}$ \\
\hline
$C_{3v}$
& $\mathcal L^{(x)}_{zx}-\mathcal L^{(y)}_{yz}$\phantom{;\nphantom{;}\phantom{+} $\mathcal L^{(x)}_{yz}+\mathcal L^{(y)}_{zx}$;\nphantom{;}\phantom{+} $\mathcal L^{(z)}_{xy}$}
& $\mathcal Q$;\nphantom{;}\phantom{+} \phantom{$\mathcal S$;}\nphantom{;}\phantom{+} $\mathcal A^{(x)}_{zx}+\mathcal A^{(y)}_{yz}$;\nphantom{;}\phantom{+} \phantom{$\mathcal A^{(x)}_{yz}-\mathcal A^{(y)}_{zx}$;}\nphantom{;}\phantom{+} \phantom{$\mathcal A^{(z)}_{xy}$;}\nphantom{;}\phantom{+} $\mathcal A^{(z)}_{xx}+\mathcal A^{(z)}_{yy}$ \\
\hline
$C_{3h}$
&
& $\mathcal Q$;\nphantom{;}\phantom{+} $\mathcal S$\phantom{;\nphantom{;}\phantom{+} $\mathcal A^{(x)}_{zx}+\mathcal A^{(y)}_{yz}$;\nphantom{;}\phantom{+} $\mathcal A^{(x)}_{yz}-\mathcal A^{(y)}_{zx}$;\nphantom{;}\phantom{+} $\mathcal A^{(z)}_{xy}$;\nphantom{;}\phantom{+} $\mathcal A^{(z)}_{xx}+\mathcal A^{(z)}_{yy}$} \\
\hline
$D_3$
& \phantom{$\mathcal L^{(x)}_{zx}-\mathcal L^{(y)}_{yz}$;\nphantom{;}\phantom{+}} $\mathcal L^{(x)}_{yz}+\mathcal L^{(y)}_{zx}$;\nphantom{;}\phantom{+} $\mathcal L^{(z)}_{xy}$
& $\mathcal Q$;\nphantom{;}\phantom{+} \phantom{$\mathcal S$;\nphantom{;}\phantom{+} $\mathcal A^{(x)}_{zx}+\mathcal A^{(y)}_{yz}$;}\nphantom{;}\phantom{+} $\mathcal A^{(x)}_{yz}-\mathcal A^{(y)}_{zx}$\phantom{;\nphantom{;}\phantom{+} $\mathcal A^{(z)}_{xy}$;\nphantom{;}\phantom{+} $\mathcal A^{(z)}_{xx}+\mathcal A^{(z)}_{yy}$} \\
\hline
$D_{3h}$
& 
& $\mathcal Q$\phantom{;\nphantom{;}\phantom{+} $\mathcal S$;\nphantom{;}\phantom{+} $\mathcal A^{(x)}_{zx}+\mathcal A^{(y)}_{yz}$;\nphantom{;}\phantom{+} $\mathcal A^{(x)}_{yz}-\mathcal A^{(y)}_{zx}$;\nphantom{;}\phantom{+} $\mathcal A^{(z)}_{xy}$;\nphantom{;}\phantom{+} $\mathcal A^{(z)}_{xx}+\mathcal A^{(z)}_{yy}$} \\
\hline
$C_n$, $n>3$
& $\mathcal L^{(x)}_{zx}-\mathcal L^{(y)}_{yz}$;\nphantom{;}\phantom{+} $\mathcal L^{(x)}_{yz}+\mathcal L^{(y)}_{zx}$;\nphantom{;}\phantom{+} $\mathcal L^{(z)}_{xy}$
& \phantom{$\mathcal Q$;\nphantom{;}\phantom{+} $\mathcal S$;\nphantom{;}\phantom{+}} $\mathcal A^{(x)}_{zx}+\mathcal A^{(y)}_{yz}$;\nphantom{;}\phantom{+} $\mathcal A^{(x)}_{yz}-\mathcal A^{(y)}_{zx}$;\phantom{\nphantom{;}\phantom{+} $\mathcal A^{(z)}_{xy}$;\nphantom{;}\phantom{+}} $\mathcal A^{(z)}_{xx}+\mathcal A^{(z)}_{yy}$ \\
\hline
$C_{nv}$, $n>3$
& $\mathcal L^{(x)}_{zx}-\mathcal L^{(y)}_{yz}$\phantom{;\nphantom{;}\phantom{+} $\mathcal L^{(x)}_{yz}+\mathcal L^{(y)}_{zx}$;\nphantom{;}\phantom{+} $\mathcal L^{(z)}_{xy}$}
& \phantom{$\mathcal Q$;\nphantom{;}\phantom{+} $\mathcal S$;\nphantom{;}\phantom{+}} $\mathcal A^{(x)}_{zx}+\mathcal A^{(y)}_{yz}$;\phantom{\nphantom{;}\phantom{+} $\mathcal A^{(x)}_{yz}-\mathcal A^{(y)}_{zx}$;\nphantom{;}\phantom{+} $\mathcal A^{(z)}_{xy}$;\nphantom{;}\phantom{+}} $\mathcal A^{(z)}_{xx}+\mathcal A^{(z)}_{yy}$ \\
\hline
$D_n$, $n>3$
& \phantom{$\mathcal L^{(x)}_{zx}-\mathcal L^{(y)}_{yz}$;\nphantom{;}\phantom{+}} $\mathcal L^{(x)}_{yz}+\mathcal L^{(y)}_{zx}$;\nphantom{;}\phantom{+} $\mathcal L^{(z)}_{xy}$
& \phantom{$\mathcal Q$;\nphantom{;}\phantom{+} $\mathcal S$;\nphantom{;}\phantom{+}} \phantom{$\mathcal A^{(x)}_{zx}+\mathcal A^{(y)}_{yz}$;\nphantom{;}\phantom{+}} $\mathcal A^{(x)}_{yz}-\mathcal A^{(y)}_{zx}$\phantom{;\nphantom{;}\phantom{+} $\mathcal A^{(z)}_{xy}$;\nphantom{;}\phantom{+} $\mathcal A^{(z)}_{xx}+\mathcal A^{(z)}_{yy}$} \\
\hline
$S_4$
& $\mathcal L^{(x)}_{zx}+\mathcal L^{(y)}_{yz}$;\nphantom{;}\phantom{+} $\mathcal L^{(x)}_{yz}-\mathcal L^{(y)}_{zx}$\phantom{+ $\mathcal L^{(z)}_{xy}$}
& \phantom{$\mathcal Q$;\nphantom{;}\phantom{+} $\mathcal S$;\nphantom{;}\phantom{+}} $\mathcal A^{(x)}_{zx}-\mathcal A^{(y)}_{yz}$;\nphantom{;}\phantom{+} $\mathcal A^{(x)}_{yz}+\mathcal A^{(y)}_{zx}$;\nphantom{;}\phantom{+} $\mathcal A^{(z)}_{xy}$;\nphantom{;}\phantom{+} $\mathcal A^{(z)}_{xx}-\mathcal A^{(z)}_{yy}$ \\
\hline
$T$
& \phantom{$\mathcal L^{(x)}_{zx}-\mathcal L^{(y)}_{yz}$;}\nphantom{;}\phantom{+} $\mathcal L^{(x)}_{yz}+\mathcal L^{(y)}_{zx}+\mathcal L^{(z)}_{xy}$
& \phantom{$\mathcal Q$;\nphantom{;}\phantom{+} $\mathcal S$;\nphantom{;}\phantom{+}} \phantom{$\mathcal A^{(x)}_{zx}+\mathcal A^{(y)}_{yz}$;\nphantom{;}\phantom{+}} $\mathcal A^{(x)}_{yz}+\mathcal A^{(y)}_{zx}+\mathcal A^{(z)}_{xy}$\phantom{;\nphantom{;}\phantom{+} $\mathcal A^{(z)}_{xx}+\mathcal A^{(z)}_{yy}$} \\
\hline
$T_d$ 
& 
& \phantom{$\mathcal Q$;\nphantom{;}\phantom{+} $\mathcal S$;\nphantom{;}\phantom{+}} \phantom{$\mathcal A^{(x)}_{zx}+\mathcal A^{(y)}_{yz}$;\nphantom{;}\phantom{+}} $\mathcal A^{(x)}_{yz}+\mathcal A^{(y)}_{zx}+\mathcal A^{(z)}_{xy}$\phantom{;\nphantom{;}\phantom{+} $\mathcal A^{(z)}_{xx}+\mathcal A^{(z)}_{yy}$} \\
\hline
$O$
& \phantom{$\mathcal L^{(x)}_{zx}-\mathcal L^{(y)}_{yz}$;}\nphantom{;}\phantom{+} $\mathcal L^{(x)}_{yz}+\mathcal L^{(y)}_{zx}+\mathcal L^{(z)}_{xy}$
&  \\
\hline
\end{tabular}
}
\end{center}
\caption{\label{table}Classification of LI-type and non-LI-type \vspace{-0.58ex}terms in~$w_{\text{ch}}$ allowed by point group symmetries. Here, $\mathcal A^{(k)}_{ij}=\Theta^{(k)}_{ij}, \Phi^{(k)}_{ij}$ and we use the notations $\mathcal Q=\mathcal A^{(x)}_{xx}-\mathcal A^{(x)}_{yy}-2\mathcal A^{(y)}_{xy}$, $\mathcal S=\mathcal A^{(y)}_{yy}-\mathcal A^{(y)}_{xx}-2\mathcal A^{(x)}_{xy}$. For the classes $C_{2h}$ ($D_{1d}$), $D_{2h}$, $D_{3d}$, $S_2$, $S_6$, $T_h$, $O_h$, and $C_{nh}$, $D_{nh}$ with $n>3$, the chiral energy density vanishes identically. Let us give an example of how to use this table. Consider a 2D system of the class $C_{\infty v}$. According to the row 11, terms with the symmetry of LIs enter $w_{\text{ch}}$ as a\vspace{-0.32ex} combination $\mathcal L^{(x)}_{zx}-\mathcal L^{(y)}_{yz}$. This\vspace{-0.32ex} corresponds to the first term on the right hand side of Eq.~(\ref{DMI_energy_linear_novel}). The combination $\mathcal A^{(x)}_{zx}+\mathcal A^{(y)}_{yz}$ with $\mathcal A=\Theta$ corresponds \vspace{-0.32ex}to the second term there, while $\mathcal A^{(x)}_{zx}+\mathcal A^{(y)}_{yz}$ with $\mathcal A=\Phi$ should be disregarded due to $\partial/\partial\phi\equiv 0$ for~$C_{\infty v}$. In 2D, $\mathcal A^{(z)}_{xx}+\mathcal A^{(z)}_{yy}$ with $\mathcal A=\Theta, \Phi$ vanish since $\nabla_z\equiv 0$.}
\end{table*}
\endgroup

Normally, for thin magnetic films and interfaces, it is assumed that $D_{\text{\tiny{LLG}}}$ is a DMI constant which is independent of $\bb n$ and defines the DMI energy density as $D_{\text{\tiny{LLG}}}\,\bb n \cdot [[\bb e_z\times\bb\nabla]\times\bb n]$~\cite{Di2015D_LLG}. According to Ref.~\cite{Hals2017}, for a 2D system of the $C_{\infty v}$ class, $D_{\text{\tiny{LLG}}}$ in Eq.~(\ref{frequency}) should, in fact, coincide with $D_{\text{as}}$ given by Eq.~(\ref{D_diff_and_D_as_def}). However, once the dependence of the functions $D_\parallel$ and $D_\perp$ on the vector~$\bb n$ is taken into account, the equality $D_{\text{\tiny{LLG}}}=D_{\text{as}}$ should also be revised. As one can see from Eq.~(\ref{D_LLG}), the result for $D_{\text{\tiny{LLG}}}$ is different from $D_{\text{as}}$ by the term $n_z\left(\partial D_\parallel/\partial n_z\right)$. Interestingly, for the model of Eq.~(\ref{Hamiltonian}), its expansion in powers of $\alpha_{\text{\tiny R}}$ starts with $\alpha_{\text{\tiny R}}^3$~\cite{supplementary}. Therefore, one could anticipate the effects of this term to be more pronounced than those of $D_{\text{diff}}$.

In Fig.~\ref{fig::D_LLG}, we plot $D_{\text{\tiny{LLG}}}$ as a function of the polar angle $\theta$ of magnetization direction, for the Bychkov-Rashba model. It is very clear that the Fourier harmonic $\cos{2\theta}$ is non-negligible, already for small values of $m\alpha_{\text{\tiny R}}^2/\Delta_{\text{sd}}$. Manifestly, in systems with strong SOC~\cite{crazySOC-1,crazySOC-2}, components of the chiral tensor can depend on $\bb n$. It would be interesting to observe such dependence experimentally. Should this happen, the proper treatment of chiral ferromagnetism must extend beyond Lifshitz invariants.

\acknowledgments
We are grateful to Rembert Duine, Olena Gomonay, Mikhail Katsnelson, and Jairo Sinova for helpful discussions. This research was supported by the JTC-FLAGERA Project GRANSPORT, the Dutch Science Foundation NWO/FOM 13PR3118, the European Research Council via Advanced Grant No. 669442, ``Insulatronics'', and by the Research Council of Norway through its Centres of Excellence funding scheme, Project No. 262633, ``QuSpin''. M.T. acknowledges the support from the Russian Science Foundation under Project 17-12-01359. 


\bibliographystyle{apsrev4-1}
\bibliography{Bib}

\begin{thebibliography}{9}
\bibitem{comment_expansions_again}
We note that, for each particular choice of $\xi(p)$ and $\zeta(p)$, one should check whether the integrals in Eqs.~(s1), (s2), (s3), (s4), (s22), and (s23) converge as well as whether the corresponding expansions actually exist
\end{thebibliography}

\supplementarystart

\centerline{\bfseries\large ONLINE SUPPLEMENTARY MATERIAL}
\vspace{6pt}
\centerline{\bfseries\large Chiral ferromagnetism beyond Lifshitz invariants}
\vspace{6pt}
\centerline{I.\,A.~Ado, A.~Qaiumzadeh, A.~Brataas, and M.~Titov}
\begin{quote}
In this Supplementary Material we provide details relevant for the text of the paper. In particular, we prove that the functions $D_\parallel$, $D_\perp$ vanish in the Bychkov-Rashba model at zero temperature if both spin subbands are partly occupied.
\end{quote}

\begin{NoHyper}

\subsection{Formal proof of the relation $D_a=D_a(n_z^2)$}
The model of Eq.~(\ref{Hamiltonian}) describes a system with rotational invariance with respect to the $z$ axis. Hence, $D_a$ cannot depend on the azimuthal angle of $\bb n$. At the same time, a simultaneous change $\theta\to-\theta$ and $\varphi\to\varphi+\pi$ does not alter the result of integration in Eq.~(\ref{DMI_tensor}). Thus, $D_\parallel$ and $D_\perp$ are even functions of $\theta$ with a period equal to~$\pi$. Since, for Fourier harmonics, $\cos{2n\theta}=F_n(\cos^2{\theta})$, we observe that, indeed, $D_a=D_a(n_z^2)$.

\subsection{$D_\parallel$ and $D_\perp$: leading-order results for small $\Delta_{\text{sd}}$}
In this section we assume $\zeta(p)>0$. Expansion of the integrands in Eq.~(\ref{D's_general}) with respect to small $\Delta_{\text{sd}}$ results in the following general expressions for the corresponding leading-order asymptotics of the functions $D_{\parallel}$ and $D_{\perp}$:
\begin{gather}
\label{Dpar_small_J_0}
D_\parallel=-\frac{\Delta^2_{\text{sd}}\sign{\alpha_{\text{\tiny R}}}}{8\pi\hbar\,\alpha^2_{\text{\tiny R}}}T
\int_{0}^{\infty}{p\,d p\,\frac{p\,\zeta'(p)+2\zeta(p)}{[p\,\zeta(p)]^3}\left(\tilde{g}_- - \tilde{g}_+\right)}
-
\frac{\Delta^2_{\text{sd}}}{8\pi\hbar\,\alpha_{\text{\tiny R}}}T
\int_{0}^{\infty}{p\,d p\,\frac{p\,\zeta'(p)+2\zeta(p)}{[p\,\zeta(p)]^2}\left(\tilde{g}'_- + \tilde{g}'_+\right)},
\\
\nonumber
D_\perp=-\frac{\Delta^2_{\text{sd}}\sign{\alpha_{\text{\tiny R}}}}{8\pi\hbar\,\alpha^2_{\text{\tiny R}}}T
\int_{0}^{\infty}{p\,d p\,\frac{p\,\zeta'(p)-\zeta(p)}{[p\,\zeta(p)]^3}\left(\tilde{g}_- - \tilde{g}_+\right)}
-
\frac{\Delta^2_{\text{sd}}}{8\pi\hbar\,\alpha_{\text{\tiny R}}}T
\int_{0}^{\infty}{p\,d p\,\frac{p\,\zeta'(p)-\zeta(p)}{[p\,\zeta(p)]^2}\left(\tilde{g}'_- + \tilde{g}'_+\right)}
\\
\label{Dperp_small_J_0}
+\frac{\Delta^2_{\text{sd}}\sign{\alpha_{\text{\tiny R}}}}{8\pi\hbar}T
\int_{0}^{\infty}{d p\,\left(\tilde{g}''_- - \tilde{g}''_+\right)},
\end{gather}
where $\tilde{g}_\pm^{(n)}=\partial^n \tilde{g}_{\pm}/\partial \xi^n$ and $\tilde{g}_{\pm}=g\left(\xi(p)\pm \vert\alpha_{\text{\tiny R}}\vert p\,\zeta(p)\right)$. Using the relation $\partial\tilde{g}_{\pm}/\partial \xi=\pm[p\,\zeta(p)]^{-1}\partial \tilde{g}_{\pm}/\partial\vert\alpha_{\text{\tiny R}}\vert$, one can also rewrite Eqs.~(\ref{Dpar_small_J_0}),~(\ref{Dperp_small_J_0}) in more compact forms
\begin{gather}
\label{Dpar_small_J}
D_\parallel=\frac{\Delta^2_{\text{sd}}\sign{\alpha_{\text{\tiny R}}}}{8\pi\hbar}T\frac{\partial}{\partial\vert\alpha_{\text{\tiny R}}\vert}\left[
\int_{0}^{\infty}{d p\,\frac{p\,\zeta'(p)+2\zeta(p)}{\vert\alpha_{\text{\tiny R}}\vert p^2\,\zeta(p)^3}\left(\tilde{g}_- - \tilde{g}_+\right)}\right],
\\
\label{Dperp_small_J}
D_\perp=\frac{\Delta^2_{\text{sd}}\sign{\alpha_{\text{\tiny R}}}}{8\pi\hbar}T\frac{\partial}{\partial \vert\alpha_{\text{\tiny R}}\vert}\left[
\int_{0}^{\infty}{d p\,\frac{p\,\zeta'(p)-\zeta(p)}{\vert\alpha_{\text{\tiny R}}\vert p^2\,\zeta(p)^3}\left(\tilde{g}_- - \tilde{g}_+\right)}\right]
+\frac{\Delta^2_{\text{sd}}\sign{\alpha_{\text{\tiny R}}}}{8\pi\hbar}T
\int_{0}^{\infty}{d p\,\left(\tilde{g}''_- - \tilde{g}''_+\right)}.
\end{gather}

It is instructive to apply the above general expressions~\cite{comment_expansions_again} to two paradigmatic models: the model of massive Dirac fermions (DF) and the Bychkov-Rashba model (BR). By setting $\xi(p)\equiv0$ and $\zeta(p)\equiv 1$ in Eq.~(\ref{Hamiltonian}), we get for the former model at $T=0$:
\be
\text{(DF):}\quad
\label{D_small_J_Dirac}
D_\parallel=
\frac{\Delta_{\text{sd}}^2}{4\pi\hbar\,\alpha_{\text{\tiny{R}}}}
\begin{cases}
\phantom{-}1, &\mu<0\\
-1, &\mu>0
\end{cases}
,
\qquad
D_\perp=0,
\e
where we used that $T g(\varepsilon)\to(\mu-\varepsilon)H\,(\mu-\varepsilon)$ and $T \partial g/\partial\varepsilon\to-\delta\,(\varepsilon-\mu)$ when temperature approaches zero (the notations $H$ and $\delta$ refer here to the Heaviside step function and the Dirac delta function, respectively). In fact, the result for $D_{\text{as}}=D_\parallel/2$ that follows from Eq.~(\ref{D_small_J_Dirac}) almost coincides with the non-perturbative one~\cite{Tserkovnyak, Koretsune, Wakatsuki}. The only difference is the absence of the band gap signature in Eq.~(\ref{D_small_J_Dirac}).

For the Bychkov-Rashba model, $\xi(p)=p^2/2m$ and $\zeta(p)\equiv 1$, we obtain
\be
\text{(BR):}\quad
\label{D_small_J_BR}
D_\parallel=
\frac{\Delta_{\text{sd}}^2}{2\pi\hbar\,\alpha_{\text{\tiny{R}}}}
\begin{cases}
\sqrt{1+2\mu/(m\alpha_{\text{\tiny{R}}}^2)},
&\mu<0\\
0, &\mu>0
\end{cases}
,
\qquad
D_\perp=
-\frac{\Delta_{\text{sd}}^2}{2\pi\hbar\,\alpha_{\text{\tiny{R}}}}\frac{\mu}{m\alpha_{\text{\tiny{R}}}^2}
\begin{cases}
1/\sqrt{1+2\mu/(m\alpha_{\text{\tiny{R}}}^2)},
&\mu<0\\
0, &\mu>0
\end{cases}
,
\e
where temperature is again set to zero. It is interesting to observe that application of a formal limit $m\to\infty$ in Eq.~(\ref{D_small_J_BR}) leads to the result
\be
\text{(BR$\vert_{m\to\infty}$):}\quad
D_\parallel=
\frac{\Delta_{\text{sd}}^2}{2\pi\hbar\,\alpha_{\text{\tiny{R}}}}
\begin{cases}
1,
&\mu<0\\
0, &\mu>0
\end{cases}
,
\qquad
D_\perp=0,
\e
which does not coincide with that of Eq.~(\ref{D_small_J_Dirac}).

Evidently, the two coefficients of the chiral tensor are not equal to each other, $D_\parallel\neq D_\perp$, in the leading order with respect to small $\Delta_{\text{sd}}$. At the same time, no dependence on $\bb n$ is present in this case (as we have stated in the main text of the paper). Hence, $D_{\text{diff}}=0$ and $D_{\text{\tiny{LLG}}}=D_{\text{as}}$, up to the order $\Delta_{\text{sd}}^2$.

\subsection{Vanishing of $D_\parallel$, $D_\perp$ in the Bychkov-Rashba model when both spin subbands are partly occupied}
For considerations of this section, it is useful to introduce the ``magnetization'' vector $\bb M=\varsigma\Delta_{\text{sd}}\bb n$ with the components $M_\perp=M_z=\varsigma\Delta_{\text{sd}}\cos{\theta}$ and $M_\parallel=\varsigma(M_x^2+M_y^2)^{1/2}=\varsigma\Delta_{\text{sd}}\sin{\theta}$. With the help of the latter, under the condition $\zeta(p)\equiv 1$, we obtain from Eqs.~(\ref{D's_general}),~(\ref{D's_micro_2}),~(\ref{D's_micro_3}) and the definition of $\Delta(\bb p)$:
\begin{gather}
\label{DparBRgen}
D_\parallel=\frac{\alpha_{\text{\tiny R}}\Delta_{\text{sd}}^2 T}{2\hbar}\int{\frac{d^2 p}{(2\pi)^2}\left(\frac{g_+-g_-}{[\Delta(\bb p)]^3}-\frac{g_+' + g_-'}{[\Delta(\bb p)]^2}\right)},
\\
\label{DperpBRgen}
D_\perp=\frac{\alpha_{\text{\tiny R}}\Delta_{\text{sd}}^2 T}{2\hbar}\int{\frac{d^2 p}{(2\pi)^2}\left(\frac{M_\parallel-\alpha_{\text{\tiny R}}p\sin{\varphi}}{M_\parallel}\right)\left(\frac{g_+-g_-}{[\Delta(\bb p)]^3}-\frac{g_+' + g_-'}{[\Delta(\bb p)]^2}\right)},
\\
\Delta(\bb p)=\sqrt{M_\parallel^2+M_\perp^2+\alpha_{\text{\tiny R}}^2 p^2-2\alpha_{\text{\tiny R}}M_\parallel p\sin{\varphi}},
\end{gather}
where one can regard $g_\pm'$ as $\partial g_\pm/\partial \xi$.
Then, using the relations
\be
\frac{\partial \Delta(\bb p)}{\partial M_\perp}=\frac{M_\perp}{\Delta(\bb p)}, \qquad
\frac{\partial \Delta(\bb p)}{\partial M_\parallel}=\frac{M_\parallel-\alpha_{\text{\tiny R}}p\sin{\varphi}}{\Delta(\bb p)}, \qquad
g_\pm'=\pm\frac{\partial g_\pm}{\partial M_\perp}\frac{\Delta(\bb p)}{M_\perp}, \qquad
g_\pm'=\pm\frac{\partial g_\pm}{\partial M_\parallel}\frac{\Delta(\bb p)}{M_\parallel-\alpha_{\text{\tiny R}}p\sin{\varphi}},
\e
it is possible to bring the integral expressions of Eqs.~(\ref{DparBRgen}),~(\ref{DperpBRgen}) to the concise forms
\be
\label{DBRgen_diff}
D_\parallel=\frac{\alpha_{\text{\tiny R}}\Delta_{\text{sd}}^2 T}{2\hbar M_\perp}\frac{\partial}{\partial M_\perp}\int{\frac{d^2 p}{(2\pi)^2}\frac{g_--g_+}{\Delta(\bb p)}}, \qquad
D_\perp=\frac{\alpha_{\text{\tiny R}}\Delta_{\text{sd}}^2 T}{2\hbar M_\parallel}\frac{\partial}{\partial M_\parallel}\int{\frac{d^2 p}{(2\pi)^2}\frac{g_--g_+}{\Delta(\bb p)}}.
\e

Next, we differentiate the above results over the chemical potential to obtain
\be
\label{dD_over_dmu}
\frac{\partial D_\parallel}{\partial\mu}=\frac{\alpha_{\text{\tiny R}}\Delta_{\text{sd}}^2}{2\hbar M_\perp}\frac{\partial}{\partial M_\perp}\int{\frac{d^2 p}{(2\pi)^2}\frac{f_--f_+}{\Delta(\bb p)}}, \qquad
\frac{\partial D_\perp}{\partial\mu}=\frac{\alpha_{\text{\tiny R}}\Delta_{\text{sd}}^2}{2\hbar M_\parallel}\frac{\partial}{\partial M_\parallel}\int{\frac{d^2 p}{(2\pi)^2}\frac{f_--f_+}{\Delta(\bb p)}},
\e
where $f_\pm=T\partial g_\pm/\partial\mu$ is expressed in terms of the Fermi-Dirac distribution as $f_\pm=\left(1+\exp{\left[(\varepsilon_\pm(\bb p)-\mu)/T\right]}\right)^{-1}$. Comparison of Eq.~(\ref{dD_over_dmu}) with Eq.~(E15) of Ref.~\cite{Ado} shows that
\be
\frac{\partial D_\parallel}{\partial\mu}=-\frac{\alpha_{\text{\tiny R}}\Delta_{\text{sd}}^2\hbar}{A M_\perp}\frac{\partial}{\partial M_\perp}\frac{\delta S_z}{M_\perp}, \qquad
\frac{\partial D_\perp}{\partial\mu}=-\frac{\alpha_{\text{\tiny R}}\Delta_{\text{sd}}^2\hbar}{A M_\parallel}\frac{\partial}{\partial M_\parallel}\frac{\delta S_z}{M_\perp},
\e
where $\delta S_z$ is the $z$-component of the total spin of conduction electrons in a unit cell of the area $A$. For the Bychkov-Rashba model, it was demonstrated~\cite{Ado} that, at zero temperature, $\delta S_z/M_\perp=-m A/(2\pi\hbar^2)$, at least as long as $\mu>\Delta_{\text{sd}}$. From this, it immediately follows that $\partial D_\parallel/\partial\mu=\partial D_\perp/\partial\mu=0$, for such values of $\mu$. Therefore, for $T=0$, the functions $D_\parallel$ and $D_\perp$ do not depend on the chemical potential, once the latter exceeds $\Delta_{\text{sd}}$.

We now temporarily restrict the analysis to the case $\mu>\Delta_{\text{sd}}$ and return to the general situation later. We also assume that temperature is set to zero. In this case, $T g_\pm$ in Eq.~(\ref{DBRgen_diff}) should be replaced with $(\mu-\varepsilon_\pm(\bb p))f_\pm$. Assuming $\xi(p)=p^2/2m$, we get
\be
D_{\parallel,\perp}=\frac{\alpha_{\text{\tiny R}}\Delta_{\text{sd}}^2}{2\hbar M_{\perp,\parallel}}\frac{\partial}{\partial M_{\perp,\parallel}}\left[
\mu\int{\frac{d^2 p}{(2\pi)^2}\frac{f_--f_+}{\Delta(\bb p)}}-\int{\frac{d^2 p}{(2\pi)^2}\frac{p^2}{2m}\frac{f_--f_+}{\Delta(\bb p)}}+\int{\frac{d^2 p}{(2\pi)^2}\left(f_-+f_+\right)}\right].
\e
The first integral inside the brackets is again proportional to $\delta S_z/M_\perp=-m A/(2\pi\hbar^2)$ and vanishes after differentiation over $M_{\perp,\parallel}$. The third integral is the density of states, which, for $\mu>\Delta_{\text{sd}}$, is also independent of $M_{\perp,\parallel}$, as was shown in Ref.~\cite{Kim}. Thus, we are left with
\be
D_{\parallel,\perp}=-\frac{\alpha_{\text{\tiny R}}\Delta_{\text{sd}}^2}{16\pi^2\hbar \,m M_{\perp,\parallel}}\frac{\partial}{\partial M_{\perp,\parallel}}\int\limits_0^{2\pi}{d\varphi\,\int\limits_{p_+}^{p_-}{dp\,\frac{p^3}{\Delta(\bb p)}}},
\e
where $p_\pm$ are the angle dependent Fermi momenta $p_{\pm}$ corresponding to $\varepsilon_{\pm}(\bb p)$ branches. Performing the integration over $p$ and utilizing the asymptotics
\be
\mu\to+\infty:\qquad p_\pm=\sqrt{2m\mu}\mp m\vert\alpha_{\text{\tiny R}}\vert + \sqrt{\frac{m}{8\mu}}\left(m\alpha_{\text{\tiny R}}^2\pm 2M_\parallel\sign{\alpha_{\text{\tiny R}}}\sin{\varphi}\right)\mp\frac{M_\perp^2+M_\parallel^2\cos{\varphi}^2}{4\vert\alpha_{\text{\tiny R}}\vert\mu}+\mathcal O(\mu^{-3/2}),
\e
we arrive at the relation
\be
D_{\parallel,\perp}=-\frac{\alpha_{\text{\tiny R}}\Delta_{\text{sd}}^2}{16\pi^2\hbar \,m M_{\perp,\parallel}}\frac{\partial}{\partial M_{\perp,\parallel}}\left[\frac{8\pi m^2}{3}\left(3\mu+2m\alpha_{\text{\tiny R}}^2\right)\right]+\mathcal O(\mu^{-1/2}),
\e
which ultimately means that both coefficients of the chiral tensor vanish at $\mu=+\infty$. At the same time, as we have already learned, $D_\parallel$ and $D_\perp$ are independent of $\mu$ when $\mu>\Delta_{\text{sd}}$. Therefore, we conclude:
\be
\label{BRdead}
D_\parallel\equiv D_\perp\equiv 0,\qquad \text{if $\mu>\Delta_{\text{sd}}$ and $T=0$.}
\e

It is easy to generalize the result~(\ref{BRdead}) and show that, in fact, $w_{\text{ch}}$ vanishes at zero temperature for all values of~$\mu$ corresponding to the case of two partly occupied subbands, i.\,e., for $\mu>\min{\varepsilon_+(\bb p)}$ (note that $\Delta_{\text{sd}}\geq\min{\varepsilon_+(\bb p)}$). The proof presented so far is based on the results of Refs.~\cite{Ado} and~\cite{Kim} established under the assumption $\mu>\Delta_{\text{sd}}$. The role of this assumption was to ensure non-negativity of the discriminant $\Delta_R$ of the cubic function $R$ defined in Eq.~(B15) of Ref.~\cite{Kim}. However, up to a positive prefactor, $\Delta_R$ coincides with $\Delta_Q(1)$, where $\Delta_Q\left(\sin^2{\phi}\right)$ is the discriminant of the quartic function $Q(p)=(\varepsilon_+(\bb p)-\mu)(\varepsilon_-(\bb p)-\mu)$ of the absolute value of momentum. And since $\mu>\min{\varepsilon_+(\bb p)}$ is only possible if the equation $Q(p)=0$ has four real solutions for $p$ when $\sin^2{\phi}=1$, we deduce
\be
\mu>\min{\varepsilon_+(\bb p)}\Rightarrow\sign{\Delta_Q(1)}=\sign{\Delta_R}\geq 0
\e
Hence, finally, for the Bychkov-Rashba model,
\be
D_\parallel\equiv D_\perp\equiv 0,\qquad \text{if $\mu>\min{\varepsilon_+(\bb p)}$ and $T=0$.}
\e

\subsection{$D_{\text{diff}}$: leading-order results for weak SOC and for small $\Delta_{\text{sd}}$ (general formulas)}
We expand the integrands in Eq.~(\ref{D's_general}) with respect to small $\alpha_{\text{\tiny R}}$, up to the fifth order. Subsequent integration over~$\varphi$ nullifies all linear and cubic contributions to $D_{\text{diff}}$, resulting in~\cite{comment_expansions_again}
\begin{multline}
\label{Ddiff_small_alpha}
D_{\text{diff}}=-
\frac{\alpha_{\text{\tiny R}}^5\sin{2\theta}}{256\pi\hbar}T\int\limits_0^{\infty}dp\,p^5\zeta^4(p)\biggl\{
\frac{105\left[3p\,\zeta'(p)+2\zeta(p)\right]}{\Delta_{\text{sd}}^5}\left(\hat{g}_--\hat{g}_+\right)+
\frac{105\left[3p\,\zeta'(p)+2\zeta(p)\right]}{\Delta_{\text{sd}}^4}\left(\hat{g}'_-+\hat{g}'_+\right)\\+
\frac{10\left[14p\,\zeta'(p)+9\zeta(p)\right]}{\Delta_{\text{sd}}^3}\left(\hat{g}''_--\hat{g}''_+\right)+
\frac{5\left[7p\,\zeta'(p)+4\zeta(p)\right]}{\Delta_{\text{sd}}^2}\left(\hat{g}'''_-+\hat{g}'''_+\right)\\+
\frac{5p\,\zeta'(p)+2\zeta(p)}{\Delta_{\text{sd}}}\left(\hat{g}^{(4)}_--\hat{g}^{(4)}_+\right)+
\frac{p\,\zeta'(p)}{3}\left(\hat{g}^{(5)}_-+\hat{g}^{(5)}_+\right)
\biggr\},
\end{multline}
where $\hat{g}_\pm=g(\xi(p)\pm\Delta_{\text{sd}})$ and $\hat{g}_\pm^{(n)}=\partial^n \hat{g}_{\pm}/\partial \xi^n$.

Similarly, expansion up to $\Delta_{\text{sd}}^4$ leads to~\cite{comment_expansions_again}
\be
\label{Ddiff_small_J}
D_{\text{diff}}=-
\frac{\Delta_{\text{sd}}^4\sign{\alpha_{\text{\tiny R}}}\sin{2\theta}}{128\pi\hbar}T\int\limits_0^{\infty}dp\,\biggl[
\frac{105\left(\tilde{g}_--\tilde{g}_+\right)}{\left\{\vert\alpha_{\text{\tiny R}}\vert p\,\zeta(p)\right\}^4}+
\frac{105\left(\tilde{g}'_-+\tilde{g}'_+\right)}{\left\{\vert\alpha_{\text{\tiny R}}\vert p\,\zeta(p)\right\}^3}+
\frac{45\left(\tilde{g}''_--\tilde{g}''_+\right)}{\left\{\vert\alpha_{\text{\tiny R}}\vert p\,\zeta(p)\right\}^2}+
\frac{10\left(\tilde{g}'''_-+\tilde{g}'''_+\right)}{\vert\alpha_{\text{\tiny R}}\vert p\,\zeta(p)}+
\left(\tilde{g}^{(4)}_--\tilde{g}^{(4)}_+\right)
\biggr],
\e
where $\tilde{g}_\pm^{(n)}=\partial^n \tilde{g}_{\pm}/\partial \xi^n$ and $\tilde{g}_{\pm}=g\left(\xi(p)\pm \vert\alpha_{\text{\tiny R}}\vert p\,\zeta(p)\right)$. In Eq.~(\ref{Ddiff_small_J}), we also assume $\zeta(p)>0$.

\subsection{$D_\parallel$, $D_\perp$, and $D_{\text{\tiny{LLG}}}$ in the Bychkov-Rashba model: expansion up to $\alpha_{\text{\tiny R}}^3$}
In the Bychkov-Rashba model, fixing $\mu<\Delta_{\text{sd}}$ and expanding the integrands in Eq.~(\ref{D's_general}) up to $\alpha_{\text{\tiny R}}^3$, we obtain the asymptotic expressions
\begin{align}
D_\parallel&=\frac{m\alpha_{\text{\tiny R}}\Delta_{\text{sd}}}{8\pi\hbar}\left\{1-\left(\frac{\mu}{\Delta_{\text{sd}}}\right)^2-\frac{1}{4}\frac{m\alpha_{\text{\tiny R}}^2}{\Delta_{\text{sd}}}\frac{\mu}{\Delta_{\text{sd}}}\left(1\phantom{3}+\,\left(\frac{\mu}{\Delta_{\text{sd}}}\right)^2+\left[3-5\left(\frac{\mu}{\Delta_{\text{sd}}}\right)^2\right]\cos{2\theta}\right)\right\},
\\
D_\perp&=\frac{m\alpha_{\text{\tiny R}}\Delta_{\text{sd}}}{8\pi\hbar}\left\{1-\left(\frac{\mu}{\Delta_{\text{sd}}}\right)^2-\frac{1}{4}\frac{m\alpha_{\text{\tiny R}}^2}{\Delta_{\text{sd}}}\frac{\mu}{\Delta_{\text{sd}}}\left(5-3\left(\frac{\mu}{\Delta_{\text{sd}}}\right)^2+\left[3-5\left(\frac{\mu}{\Delta_{\text{sd}}}\right)^2\right]\cos{2\theta}\right)
\right\}.
\end{align}
According to the definition of Eq.~(\ref{D_LLG}), they determine the expansion
\be
\label{D_LLG_expansion}
D_{\text{\tiny{LLG}}}=\frac{m\alpha_{\text{\tiny R}}\Delta_{\text{sd}}}{8\pi\hbar}
\left\{
1-\left(\frac{\mu}{\Delta_{\text{sd}}}\right)^2-\frac{1}{2}\frac{m\alpha_{\text{\tiny R}}^2}{\Delta_{\text{sd}}}\frac{\mu}{\Delta_{\text{sd}}}\left(3\left[1-\left(\frac{\mu}{\Delta_{\text{sd}}}\right)^2\right]+\left[3-5\left(\frac{\mu}{\Delta_{\text{sd}}}\right)^2\right]\cos{2\theta}\right)
\right\},
\e
which we use in Fig.~\ref{fig::D_LLG} of the paper. Note that, in this model, $D_\parallel-D_\perp$ is independent of $\theta$, in the order $\alpha_{\text{\tiny R}}^3$, and $D_{\text{diff}}$, indeed, vanishes.

\subsection{Chiral energy density for the classes $C_1$, $C_{1v}$, and $C_{1h}$}
In the chiral energy density, the classes $C_1$, $C_{1v}$, and $C_{1h}$ allow the following LIs:
\begin{align}
\text{$\bigl(C_1\bigr)$:}&\quad \mathcal L^{(x)}_{xy};\,\,\mathcal L^{(x)}_{yz};\,\,\mathcal L^{(x)}_{zx};\,\,\mathcal L^{(y)}_{xy};\,\,\mathcal L^{(y)}_{yz};\,\,\mathcal L^{(y)}_{zx};\,\,\mathcal L^{(z)}_{xy};\,\,\mathcal L^{(z)}_{yz};\,\,\mathcal L^{(z)}_{zx},\\
\label{C1vL}
\text{$\bigl(C_{1v}\bigr)$:}&\quad \phantom{\mathcal L^{(x)}_{xy};\,\,\mathcal L^{(x)}_{yz};\,\,}\,\mathcal L^{(x)}_{zx};\,\,\mathcal L^{(y)}_{xy};\,\,\mathcal L^{(y)}_{yz};\,\,\phantom{\mathcal L^{(y)}_{zx};\,\,\mathcal L^{(z)}_{xy};\,\,\mathcal L^{(z)}_{yz};\,\,}\,\mathcal L^{(z)}_{zx},\\
\text{$\bigl(C_{1h}\bigr)$:}&\quad \mathcal L^{(x)}_{xy};\,\,\phantom{\mathcal L^{(x)}_{yz};\,\,\mathcal L^{(x)}_{zx};\,\,}\,\mathcal L^{(y)}_{xy};\,\,\phantom{\mathcal L^{(y)}_{yz};\,\,\mathcal L^{(y)}_{zx};\,\,\mathcal L^{(z)}_{xy};\,\,}\,\mathcal L^{(z)}_{yz};\,\,\mathcal L^{(z)}_{zx},\\
\end{align}
and non-LI-type terms with the following coefficients:
\begin{align}
\text{$\bigl(C_1\bigr)$:}&\quad \mathcal A^{(x)}_{xx};\,\,\mathcal A^{(x)}_{yy};\,\,\mathcal A^{(x)}_{xy};\,\,\mathcal A^{(x)}_{yz};\,\,\mathcal A^{(x)}_{zx};\,\,\mathcal A^{(y)}_{xx};\,\,\mathcal A^{(y)}_{yy};\,\,\mathcal A^{(y)}_{xy};\,\,\mathcal A^{(y)}_{yz};\,\,\mathcal A^{(y)}_{zx};\,\,\mathcal A^{(z)}_{xx};\,\,\mathcal A^{(z)}_{yy};\,\,\mathcal A^{(z)}_{xy};\,\,\mathcal A^{(z)}_{yz};\,\,\mathcal A^{(z)}_{zx},\\
\label{C1vA}
\text{$\bigl(C_{1v}\bigr)$:}&\quad \mathcal A^{(x)}_{xx};\,\,\mathcal A^{(x)}_{yy};\phantom{\,\,\mathcal A^{(x)}_{xy};\,\,\mathcal A^{(x)}_{yz};\,\,}\,\mathcal A^{(x)}_{zx};\phantom{\,\,\mathcal A^{(y)}_{xx};\,\,\mathcal A^{(y)}_{yy};\,\,}\,\mathcal A^{(y)}_{xy};\,\,\mathcal A^{(y)}_{yz};\phantom{\,\,\mathcal A^{(y)}_{zx};\,\,}\,\mathcal A^{(z)}_{xx};\,\,\mathcal A^{(z)}_{yy};\,\,\phantom{\mathcal A^{(z)}_{xy};\,\,\mathcal A^{(z)}_{yz};\,\,}\,\mathcal A^{(z)}_{zx},\\
\text{$\bigl(C_{1h}\bigr)$:}&\quad \mathcal A^{(x)}_{xx};\,\,\mathcal A^{(x)}_{yy};\,\,\mathcal A^{(x)}_{xy};\,\,\phantom{\mathcal A^{(x)}_{yz};\,\,\mathcal A^{(x)}_{zx};\,\,}\,\mathcal A^{(y)}_{xx};\,\,\mathcal A^{(y)}_{yy};\,\,\mathcal A^{(y)}_{xy};\,\,\phantom{\mathcal A^{(y)}_{yz};\,\,\mathcal A^{(y)}_{zx};\,\,\mathcal A^{(z)}_{xx};\,\,\mathcal A^{(z)}_{yy};\,\,\mathcal A^{(z)}_{xy};\,\,}\,\mathcal A^{(z)}_{yz};\,\,\mathcal A^{(z)}_{zx},
\end{align}
where $\mathcal A^{(k)}_{ij}=\Theta^{(k)}_{ij}, \Phi^{(k)}_{ij}$.

\end{NoHyper}


\end{document}